\documentclass[prl,twocolumn]{revtex4-1}

\usepackage{amssymb,amsmath}

\usepackage{epsfig}
\usepackage{bm}
\usepackage{color}
\usepackage{graphicx}

\begin{document}

\title{Navier-Stokes hydrodynamics in a granular gas fluidized by a non-uniform stochastic field}
\author{Francisco Vega Reyes}
\affiliation{Departamento de F\'isica, Universidad de Extremadura, 06071 Badajoz, Spain}
\email{fvega@unex.es}


\begin{abstract}
We show a case of steady flow in a granular gas that, for small shear rates, is accurately described by Navier-Stokes hydrodynamics, even for high inelasticity. The (low density) granular gas is composed of identical inelastic spheres and is fluidized by a temperature field produced by a stochastic volume force plus thermal walls. We prove that the theoretical Navier-Stokes transport coefficients exhibit good agreement with numerical solutions of the corresponding the kinetic equation (direct simulation Monte Carlo method) in the steady state.
\end{abstract}

\date{\today}
\maketitle

The transport properties of low density granular matter have been studied in a number of theoretical works \cite{D01,BP04} whose final objective is to be of use in applications (granular matter is present in a variety of transport problems in  different industry sectors) \cite{G03}. Due to kinetic energy loss during collisions, the inelastic particles need to be continuously excited. In most real granular transport problems, this implies some kind of energy input  from the system boundaries, which gives rise to inhomogeneous steady flows. This may cause ultimately the hydrodynamic description fail, if the spatial gradients of average fields cease to be small compared to the typical spatial variation rate of the microscopic length scale \cite{CC70,D01,G03}. More specifically, for Couette-Fourier plane flows, it has been shown and quantified the deviation from Navier-Stokes (NS) description outside the quasi-elastic limit \cite{VU09}.

We show in this work that this failure may be concealed for granular gases fluidized by volume forces, if appropriate NS transport coefficients are used \cite{GM02}, even at strong inelasticity.  In the present work, the stochastic force is an inhomogeneous field to which the granular temperature adjusts locally. In particular, our stochastic force field will mimic a Fourier flow. This makes sense if we consider that the granular gas is fluidized by some interstitial fluid that is heated from the boundaries, that in our case are two infinite parallel walls. Moreover, this kind of forcing is of practical interest since the stochastic thermostat may resemble the fluidization produced by a turbulent air flow \cite{OLDLD04}, for instance. We will consider that this force exactly balances inelastic cooling, thus producing a steady flow with uniform heat flux (like the Fourier flow in a molecular gas) \cite{VSG10}. The steady base state in this work - to the author's knowledge-  has not been reported previously. In addition, we will also show that NS hydrodynamics still holds if over the base Fourier flow a small shear is input, this applying for a wide range of inelasticity conditions. 

This example of validity of NS hydrodynamic description for a steady granular flows at strong inelasticities  may support the idea that scale separation is not necessarily always limited in an inherent way in rapid granular flows, a point that has been debated in a number of works on granular dynamics for a long time \cite{G03}. In fact, and as we will see, NS hydrodynamics is valid here because the base state has strictly the same spatial gradients size as the Fourier state for a molecular gas \cite{CC70}. This means that this steady granular flow may be described in the frame of NS hydrodynamics no matter how strong inelasticity is, as long as the spatial gradients of the average fields are small. 

We deal with a set of inelastic smooth hard disks/spheres with a diameter $\sigma$ and a mass $m$. As usual for inelastic smooth hard spheres, inelasticity is characterized by the coefficient of normal restitution $\alpha$ \cite{BDKS98,VU09}. This coefficient ranges from 1 for purely elastic collisions to 0 for purely inelastic collisions. Collisional rules for this model may be found elsewhere \cite{G03}. The system is limited by two infinite parallel hard walls, located at planes $y=\pm L/2$ respectively. The walls act like two distinct kinetic energy sources, characterized with temperatures $T_\pm$. They may have relative movement (wall velocities $U_\pm$ respectively), eventually inducing the particles to continuously flow along the channel between them. In addition, the inelastic particles interact with a surrounding kinetic energy source. We model this interaction by means of a stochastic volume force with the form of a Gaussian white noise  \cite{MTC07,GSVP11,OLDLD04}, which we consider to be in general a function of the spatial coordinate $y$. This force, that we denote as $\mathbf{F}^\mathrm{st}(y)$, fulfills at each $y$ point the conditions

\begin{eqnarray}
& &\langle {\bf F}_i^{\text{st}}(y,t) \rangle ={\bf 0},\: \nonumber \\ 
& &\langle {\bf F}_i^{\text{st}}(y,t) {\bf F}_j^{\text{st}}(y,t') \rangle =\mathbf{1}m^2 \xi(y)^2 \delta(y,t-t'),
\label{Fst}
\end{eqnarray} being $\xi^2$ the noise intensity \cite{MTC07,GSVP11}, $\mathbf{1}$ the unit matrix in $d$ dimensional space and  $\langle A\rangle$ indicates average of  $A$. It is to be remarked that whereas in other works this stochastic force has a uniform intensity, here we consider a more general situation where the interaction of the granular gas with its surroundings is space-dependent, and thus we have $\xi(y)$. The white noise term is known to yield a Fokker-Planck-like operator in the inelastic Boltzmann equation,  

\begin{equation}
\left(\frac{\partial}{\partial t}
+\mathbf{v}\cdot\nabla\right)f(\mathbf{r},\mathbf{v};t)-\frac{1}{2}\xi(y)^2\frac{\partial^2}{\partial v^2}=J[\mathbf{v}|f,f], 
\label{BE}
\end{equation} 
where $J$ is the collisional integral for inelastic hard spheres and whose expression may be found elsewhere \cite{G03}. Taking into account the system geometries, and for a steady state, the general kinetic equation \eqref{BE} reduces to

 \begin{equation}
v_y\frac{\partial f}{\partial y}-\frac{1}{2}\xi(y)^2\frac{\partial^2f}{\partial v^2}=J[f,f].
\label{rBE}
\end{equation} 
Integrating equation \eqref{rBE} times the three first velocity moments, we obtain mass, momentum and energy balance equations. From now on, and as in \cite{VU09}, we will use a special scaled space variable $l$ that is related to $y$ through the relation $\sqrt{T/T_r}\partial /\partial y=\partial /\partial l$. Here, $T_r$ is the reference unit of temperature, that we choose to be the granular temperature at the point with its lowest value \cite{VU09} (i.e., $T_r=T_-$). We complete our set of units with $m$ for mass, $n_r$ for particle density ($n_r$ is the density at lower wall), $\lambda_r=\Lambda_3/(\sqrt{2}n_r\sigma^{2})$ for length (mean free path at lower wall) and $\nu_r^{-1}$ for time (with $\nu_r=\sqrt{2T_r/m}(n_r\sigma^{2}/\Lambda_3)$ the collision frequency at the lower wall, since $T_r=T_-$). In these definitions, $\Lambda_3=5\sqrt{2}\Gamma(3/2)/8$ \cite{VU09}. From now on, we will indicate reduced spatial coordinate as $\hat y$ and analogously for all dimensionless magnitudes, except transport coefficients. In this system, the relevant balance equations, together with NS constitutive equations yield \cite{VU09}

\begin{eqnarray}
& &\frac{\partial \hat{p}}{\partial \hat l}=0, \quad \frac{\partial\hat u_{y}}{\partial\hat l}=\frac{\hat\gamma}{\eta^*(\alpha)}\equiv a,
\label{Pij}\\
& &\frac{\partial^2 \hat T}{\partial\hat l^2}=\hat\Gamma(\alpha), 
\label{difT}
\end{eqnarray} where $\hat\gamma$ for this geometry is a constant and by definition we call $a$ the \textit{shear rate} (constant and also dimensionless \cite{VSG10}) and in \eqref{Pij} we have used that for NS hydrodynamics: a) all diagonal elements of the stress tensor are equal and thus equal to the hydrostatic pressure $p$, b) $P_{xy}=\eta(\alpha)\partial u_{ x}/\partial y$, c) $q_y=-\lambda\partial T/\partial y-\mu\partial n/\partial y $ \cite{BDKS98}, with $\eta$ the viscosity and $\lambda, \mu$ the heat flux transport coefficients (math procedures are analogous to those in \cite{VU09}, where more details can be found). With the presence of the stochastic volume force $\hat\Gamma (\alpha)$ takes the form

\begin{equation}
\hat{\Gamma}(\alpha)\equiv\frac{\phi_3}{\beta^*(\alpha)}\left(\frac{3}{2}\left(\zeta^*(\alpha)-\frac{\hat\xi(\hat l)^2}{\hat T(\hat l)^{1/2}}\right)-\eta^*(\alpha)a^2\right) ,
\label{GR}
\end{equation} where $\phi_3=6/15$, $\beta^*(\alpha)\equiv\kappa^*(\alpha)-\mu^*(\alpha)$, that we call \textit{effective thermal conductivity}.  The expressions of the reduced transport coefficients (viscosity $\eta^*(\alpha)$, thermal conductivity $\kappa^*(\alpha)$, and $\mu^*(\alpha)$), and cooling rate $\zeta^*(\alpha)$ that we used are the ones from the NS theory derived by Garz\'o and Montanero for the granular gas heated by the stochastic force (with constant $\xi$ )\cite{GM02}, by means of Chapman-Enskog theory \cite{CC70,D01,BP04}.

As shown in a previous work, Couette-Fourier flows are characterized by three independent microscopic/macroscopic relative scales (or Knudsen numbers) \cite{VU09}, which are: $a$, $|\hat\Gamma|^{1/2}$ and $\Delta\hat T/\hat L$, with $\Delta \hat T\equiv  T_+-T_-$. A way to achieve a steady flow with small gradients, is to look for flows with $a=0, \hat\Gamma=0$ and $\Delta\hat T/\hat L\lesssim 1$. Obviously, in order to achieve a steady flow we need some source of inhomogeneity and thus we cannot keep all three Knudsen numbers equal to zero. Since as we know from previous works \cite{VSG10,VU09} that $\Delta \hat T/\hat L$ is the one having the smallest effect on deviations from NS hydrodynamics, we choose precisely this number as not null for our reference state. Condition $a=0$ is simply achieved by turning off shearing from the boundaries; i.e. $\Delta U\equiv U_+-U_-=0$ and with condition $\hat\Gamma=0$ from \eqref{GR} we obtain

\begin{equation}
\hat \xi(\hat l)^2=\zeta^*(\alpha)\hat T(\hat l)^{1/2}.
\label{xiLTu}
\end{equation} Condition $\hat\Gamma=0$; i.e., $\partial^2T/\partial\hat l^2=0$ from \eqref{difT}, also implies that heat flux is uniform throughout the system \cite{VU09} and therefore our base steady state is a \textit{new} element of the LTu class of granular flows \cite{VSG10}. On the other hand, it is known that $\hat\Gamma=0$ solutions with $T_-\neq T_+$ lead to profiles of the type $\hat T(\hat l)=A\hat l +B$, where $A, B$ are constants \cite{VU09,VSG10}. 

Summarizing, we obtain profiles with the properties: i) constant hydrostatic pressure $\hat p$, ii) null normal stress differences $P_{ii}=P_{jj}$, iii) constant $q_y$ and null $q_x$, iv) linear $\hat T(\hat l)$ profiles, or equivalently, $\hat T(\hat y)^{2/3}$ is linear in $y$ \cite{VSG10}, and most importantly, v) their transport coefficients are well described by hydrodynamics at NS order, as deduced from Chapman-Enskog theory. Properties i) to iii) are fulfilled by all Couette-Fourier flows (as this one) at NS order and condition iv) is fulfilled by all uniform heat flux flows. Finally, condition v) is fulfilled by all flows with not too strong gradients. That is to say, this new type of steady flow is the first one to be reported that fulfills all conditions for NS hydrodynamics even at high inelasticities, and is an LTu flow \cite{VSG10}. 

We will prove i)-v) are fulfilled even outside the quasi-elastic limit $\alpha\simeq 1$ to inelasticities as high as $\alpha=0.5$. For this purpose, we will compare theoretical results with simulation data from Monte Carlo (DSMC) solution  \cite{PS05} of the kinetic equation \eqref{rBE}. For details on the computational methods see \cite{VU09,VSG10}. Implementation of the term coming from the stochastic force, is described for instance in \cite{MS00,GSVP11}. In the present work, the only variation is that noise intensity is space-dependent, following \eqref{xiLTu}. For the solution to be self-consistent, we have extracted simulation temperature profiles from LTu states obtained in a previous work \cite{VSG10}, introducing them in condition \eqref{xiLTu}. In the simulation data presented in all figures we have considered a system with $T_+/T_-=5$ and $L=15\overline\lambda$ ($\overline\lambda=\sqrt{2}\pi\overline n\sigma^{2}$, $\overline n$ being the average density in the system). We have checked that properties i) to iv) are indeed fulfilled in all simulations. In order check property v) and to measure possible deviations from NS behavior we have done simulations with gradually increasing rate. These states have been obtained in DSMC by using the same $\hat\xi(y)^2$ profiles as for $a=0$ base states; i.e., those fulfilling \eqref{xiLTu}, but adding a shear. Thus, from \eqref{Pij}, \eqref{GR} and \eqref{xiLTu} for our non-zero shear ($a\neq 0$) steady states 

\begin{equation}
\hat{\Gamma}(\alpha)=-\frac{\phi_3}{\beta^*(\alpha)}\eta^*(\alpha)a^2\neq 0,
\label{GRshear}
\end{equation} and obviously as we increase shear rate $a$ we gradually depart from NS hydrodynamics as both $a$ and $|\hat\Gamma|^{1/2}$ increase.

\begin{figure}
\includegraphics[height=5.1cm]{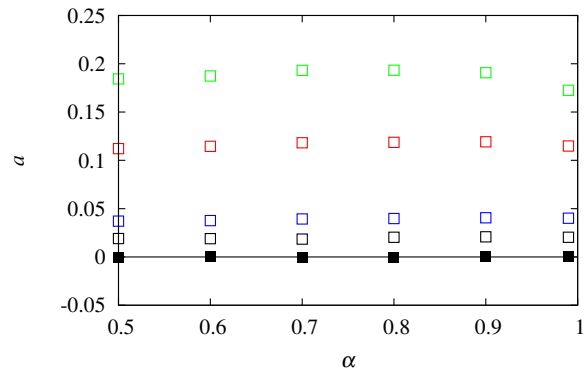}
\caption{DSMC simulation data for shear rate vs. coefficient of restitution $\alpha$, for different relative wall velocities: $\Delta \hat U=0$ with $a=0$ (black, solid), $\Delta\hat U=0.25$ with $a\simeq 0.019$ (black, open), $\Delta\hat U=1$ with $a\simeq 0.04$ (blue), $\Delta\hat U=3$ with $a\simeq 0.10$ (red) and $\Delta\hat U=5$ with $a\simeq 0.18$ (green). For this and following figures: $T_+/T_-=5$, $L=15\overline\lambda$ with $\overline\lambda=\sqrt{2}\pi\overline n\sigma^{d-1}$, $\overline n$ being the average density in the system.} 
\label{figshear}
\end{figure}

\begin{figure}
\includegraphics[height=7.25cm]{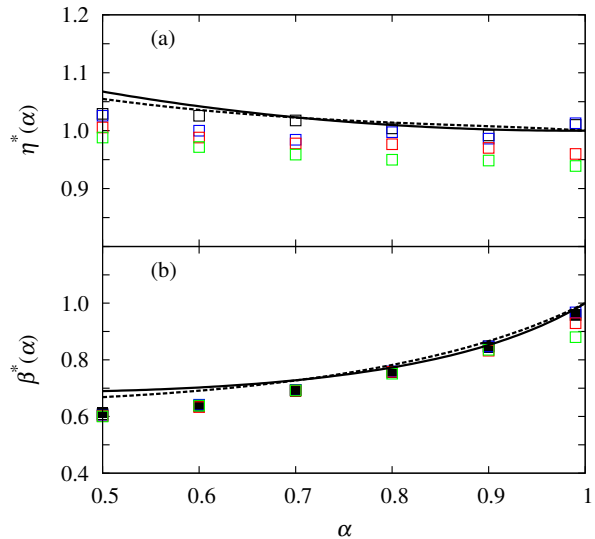}
\caption{(a) Viscosity vs. coefficient of restitution $\alpha$. (b) Effective thermal conductivity vs. $\alpha$. In both panels, lines represent the results for the theoretical NS coefficient (solid for standard Sonine approximation and dashed for improved Sonine approximation) whereas the symbols stand for DSMC simulations. Several DSMC series have been represented, with increasing relative wall velocity: $\Delta\hat U=0$ with $a=0$ (black, solid), $\Delta\hat U=0.25$ with $a\simeq 0.019$ (black, open), $\Delta \hat U=1$ with $a\simeq 0.04$ (blue), $\Delta\hat U=3$ with $a\simeq 0.10$ (red) and $\Delta\hat U=3$ with $a\simeq 0.10$ (green).} 
\label{figcoefs}
\end{figure}

Prior to comparing NS theory and simulation, it is useful to inspect the behavior of the shear rate $a$ as a function of relative wall velocity $\Delta U$, as extracted from DSMC simulation data. As we can see in Fig. \ref{figshear}, all points in constant $\Delta U$ series have approximately the same shear rate, independently of the value of the coefficient of restitution $\alpha$. This result is of help since it implies that transport coeffcient graphs vs. $\alpha$ are at approximately constant shear rate $a$ and hence constant Knudsen numbers, since $\Delta \hat T/ \hat L$ is kept constant in the simulations and from \eqref{GRshear} constant $a$ implies constant $\hat\Gamma$. We also performed an additional series with varying $\Delta\hat T/ \hat L$ (not shown in figures) and checked that the measured transport coefficients do not depend on its value, analogously to the result in a previous work on granular flows with uniform heat flux \cite{VSG10}. In Fig. \ref{figcoefs} we present the theory-simulation comparison for reduced viscosity $\eta^*(\alpha)$ and effective thermal conductivity $\beta^*(\alpha)$, for several values of relative wall velocity. For the viscosity the agreement between theory and simulation is excellent if shear rate is small but for increasing shear rates the disagreement increases. Respect to thermal effective conductivity, the agreement is very good in the full range of shear rates and inelasticities represented here. In both cases, the theory-simulation agreement in the region of high inelasticity is improved when the distribution function in the Chapman-Enskog method is not the Maxwellian but an approximation to the solution of the homogeneous state for the heated granular gas (slightly different from  Maxwellian \cite{GVM07}).

\begin{figure}
\includegraphics[height=7.25cm]{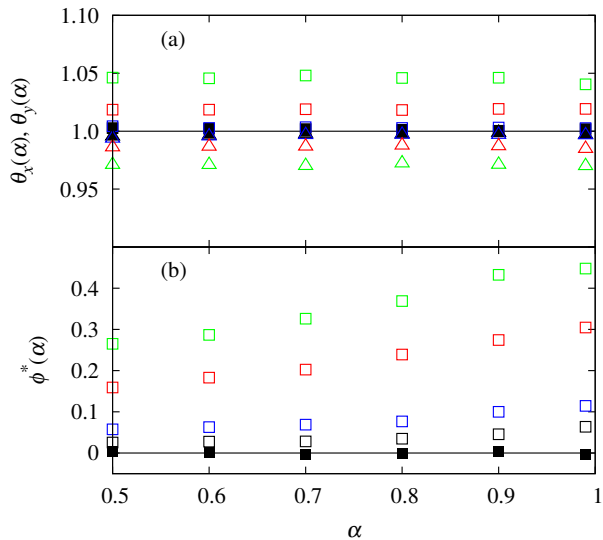}
\caption{(a) DSMC data for normal stress differences $\theta_x\equiv P_{xx}/p, \theta_y\equiv P_{yy}/p$ vs. coefficient of restitution $\alpha$. Square symbols stand for $\theta_x$ and triangles stand for $\theta_y$. (b) DSMC data for thermal cross conductivity $\phi^*$ vs.  $\alpha$. In both panels, solid symbols represent the no-shear case and open symbols are the sheared states: $\Delta\hat U=0.25$ with $a\simeq 0.019$ (black), $\Delta\hat U=1$ (blue) with $a\simeq0.04$, $\Delta\hat U=3$ with $a\simeq 0.10$ (red), $\Delta\hat U=5$ with $a\simeq 0.18$ (green).}
\label{figreol}
\end{figure}

Next we look for other possible deviations from the NS behavior in the simulations. We report in Fig. \ref{figreol}  (a) the normal stress differences $\theta_x\equiv P_{xx}/p, \theta_y\equiv P_{yy}/p$ measured in the simulations as a function of the coefficient of restitution $\alpha$, for several series at different constant shear. As we see, deviations from the NS behavior ($\theta_i$ deviations from unity) are small even at moderately high values of shear. In this case the coefficient of restitution seems not to have an important impact on the behavior of both  $\theta_x$ and $\theta_y$, consistent with the behavior observed previously in analogous LTu flows \cite{VSG13}. In Fig. \ref{figreol} (b) we represent the value of the cross conductivity coefficient, defined as $\phi^*(\alpha)=q_x/(\lambda_0\partial T/\partial y)$ \cite{VSG13}, where $q_x$ is the heat flux in the $x$ direction (non-linear effect also), and $\lambda_0$ is the heat conductivity for an elastic gas. As we can see, the deviations from linear behavior ($\phi^*(\alpha)=0$) are more significant here. However, as expected, the NS behavior is recovered for null/small shear rate (less than $\%4$ deviation respect to NS for the smallest shear rate series represented). Moreover, it is remarkable that these deviations increase as we approach the quasi-elastic limit (i.e., more inelastic flows show less significantly non-Newtonian behavior here). This is consistent with the fact that $|\hat\Gamma(\alpha)|$ increases as we approach the quasi-elastic limit \cite{VSG13}, if we are in the region $\hat\Gamma(\alpha)<0$ like in our case. A higher $|\hat\Gamma(\alpha)|$ implies higher Knudsen number in the system and thus a state further from equilibrium.


In summary, the transport properties observed in this heated granular system confirm that NS hydrodynamics applies for this system as long as the relevant Knudsen numbers in the problem do keep small. More generally, the fact that this observation is independent of the degree of inelasticity supports the hypothesis that inelasticity by itself does not cause necessarily the breakdown of scale separation in granular gases. The dynamics behavior is actually a bit more complex and the energy source necessary to fluidize the system may cancel out the gradients inherently produced by inelasticity, bringing the granular gas back to the realm of NS hydrodynamics. In fact the theory-simulation disagreement observed here for the viscosity and thermal conductivity as inelasticity increases is consistent with the failure of the Sonine approximation used in the zeroth order distribution function (and for this reason the DSMC data agree better with the improved Sonine approximation, see Fig. \ref{figcoefs}), and not with a failure of hydrodynamics.


Furthermore, Figs. 2-3 indicate a natural limit for NS hydrodynamic in this type of configuration, since the non-linear effects appear first for elastic gases in this case (in particular, see Fig. 3 (b)). Therefore, we show more evidence in this work that NS hydrodynamics is valid for granular gases just in the same way is for molecular gases: as long as the spatial gradients are not large. This result is general so we can apply NS hydrodynamics to systems and problems derived from the one in this work. For instance, to granular segregation in LTu flows \cite{VGK13}.



The author acknowledges support of the Spanish Government through Grants FIS2010-12587 (partially financed by FEDER funds and by Junta de Extremadura through Grant No. GRU10158) and MAT2009-12351-C02-02.

\bibliographystyle{apsrev}
\bibliography{nLTu}

\begin{thebibliography}{16}
\expandafter\ifx\csname natexlab\endcsname\relax\def\natexlab#1{#1}\fi
\expandafter\ifx\csname bibnamefont\endcsname\relax
  \def\bibnamefont#1{#1}\fi
\expandafter\ifx\csname bibfnamefont\endcsname\relax
  \def\bibfnamefont#1{#1}\fi
\expandafter\ifx\csname citenamefont\endcsname\relax
  \def\citenamefont#1{#1}\fi
\expandafter\ifx\csname url\endcsname\relax
  \def\url#1{\texttt{#1}}\fi
\expandafter\ifx\csname urlprefix\endcsname\relax\def\urlprefix{URL }\fi
\providecommand{\bibinfo}[2]{#2}
\providecommand{\eprint}[2][]{\url{#2}}

\bibitem[{\citenamefont{Dufty}(2001)}]{D01}
\bibinfo{author}{\bibfnamefont{J.~W.} \bibnamefont{Dufty}},
  \bibinfo{journal}{Adv. Complex Sys.} \textbf{\bibinfo{volume}{4}},
  \bibinfo{pages}{397} (\bibinfo{year}{2001}).

\bibitem[{\citenamefont{Brilliantov and P\"oschel}(2004)}]{BP04}
\bibinfo{author}{\bibfnamefont{N.~V.} \bibnamefont{Brilliantov}}
  \bibnamefont{and}
  \bibinfo{author}{\bibfnamefont{T.}~\bibnamefont{P\"oschel}},
  \emph{\bibinfo{title}{Kinetic Theory of Granular Gases}}
  (\bibinfo{publisher}{Oxford University Press, Oxford}, \bibinfo{year}{2004}).

\bibitem[{\citenamefont{Goldhirsch}(2003)}]{G03}
\bibinfo{author}{\bibfnamefont{I.}~\bibnamefont{Goldhirsch}},
  \bibinfo{journal}{Annu. Rev. Fluid Mech.} \textbf{\bibinfo{volume}{35}},
  \bibinfo{pages}{267} (\bibinfo{year}{2003}).

\bibitem[{\citenamefont{Chapman and Cowling}(1970)}]{CC70}
\bibinfo{author}{\bibfnamefont{C.}~\bibnamefont{Chapman}} \bibnamefont{and}
  \bibinfo{author}{\bibfnamefont{T.~G.} \bibnamefont{Cowling}},
  \emph{\bibinfo{title}{The Mathematical Theory of Non-Uniform Gases}}
  (\bibinfo{publisher}{Cambridge University Press, Cambridge},
  \bibinfo{year}{1970}), \bibinfo{edition}{3rd} ed.

\bibitem[{\citenamefont{{Vega Reyes} and Urbach}(2009)}]{VU09}
\bibinfo{author}{\bibfnamefont{F.}~\bibnamefont{{Vega Reyes}}}
  \bibnamefont{and} \bibinfo{author}{\bibfnamefont{J.~S.}
  \bibnamefont{Urbach}}, \bibinfo{journal}{J. Fluid Mech.}
  \textbf{\bibinfo{volume}{636}}, \bibinfo{pages}{279} (\bibinfo{year}{2009}).

\bibitem[{\citenamefont{Garz\'o and Montanero}(2002)}]{GM02}
\bibinfo{author}{\bibfnamefont{V.}~\bibnamefont{Garz\'o}} \bibnamefont{and}
  \bibinfo{author}{\bibfnamefont{J.~M.} \bibnamefont{Montanero}},
  \bibinfo{journal}{Physica A} \textbf{\bibinfo{volume}{313}},
  \bibinfo{pages}{336} (\bibinfo{year}{2002}).

\bibitem[{\citenamefont{Ojha et~al.}(2004)\citenamefont{Ojha, Lemieux, Dixon,
  Liu, and Durian}}]{OLDLD04}
\bibinfo{author}{\bibfnamefont{R.~P.} \bibnamefont{Ojha}},
  \bibinfo{author}{\bibfnamefont{P.-A.} \bibnamefont{Lemieux}},
  \bibinfo{author}{\bibfnamefont{P.~K.} \bibnamefont{Dixon}},
  \bibinfo{author}{\bibfnamefont{A.~J.} \bibnamefont{Liu}}, \bibnamefont{and}
  \bibinfo{author}{\bibfnamefont{D.~J.} \bibnamefont{Durian}},
  \bibinfo{journal}{Nature} \textbf{\bibinfo{volume}{427}},
  \bibinfo{pages}{521} (\bibinfo{year}{2004}).

\bibitem[{\citenamefont{{Vega Reyes} et~al.}(2010)\citenamefont{{Vega Reyes},
  Santos, and Garz\'o}}]{VSG10}
\bibinfo{author}{\bibfnamefont{F.}~\bibnamefont{{Vega Reyes}}},
  \bibinfo{author}{\bibfnamefont{A.}~\bibnamefont{Santos}}, \bibnamefont{and}
  \bibinfo{author}{\bibfnamefont{V.}~\bibnamefont{Garz\'o}},
  \bibinfo{journal}{Phys. Rev. Lett.} \textbf{\bibinfo{volume}{104}},
  \bibinfo{pages}{028001} (\bibinfo{year}{2010}).

\bibitem[{\citenamefont{Brey et~al.}(1998)\citenamefont{Brey, Dufty, Kim, and
  Santos}}]{BDKS98}
\bibinfo{author}{\bibfnamefont{J.~J.} \bibnamefont{Brey}},
  \bibinfo{author}{\bibfnamefont{J.~W.} \bibnamefont{Dufty}},
  \bibinfo{author}{\bibfnamefont{C.~S.} \bibnamefont{Kim}}, \bibnamefont{and}
  \bibinfo{author}{\bibfnamefont{A.}~\bibnamefont{Santos}},
  \bibinfo{journal}{Phys. Rev. E} \textbf{\bibinfo{volume}{58}},
  \bibinfo{pages}{4638} (\bibinfo{year}{1998}).

\bibitem[{\citenamefont{Marini-Bettolo-Marconi
  et~al.}(2007)\citenamefont{Marini-Bettolo-Marconi, Tarazona, and
  Cecconi}}]{MTC07}
\bibinfo{author}{\bibfnamefont{U.}~\bibnamefont{Marini-Bettolo-Marconi}},
  \bibinfo{author}{\bibfnamefont{P.}~\bibnamefont{Tarazona}}, \bibnamefont{and}
  \bibinfo{author}{\bibfnamefont{F.}~\bibnamefont{Cecconi}},
  \bibinfo{journal}{J. Chem. Phys.} \textbf{\bibinfo{volume}{126}},
  \bibinfo{pages}{164904} (\bibinfo{year}{2007}).

\bibitem[{\citenamefont{Gradenigo et~al.}(2011)\citenamefont{Gradenigo,
  Sarracino, Villamaina, and Puglisi}}]{GSVP11}
\bibinfo{author}{\bibfnamefont{G.}~\bibnamefont{Gradenigo}},
  \bibinfo{author}{\bibfnamefont{A.}~\bibnamefont{Sarracino}},
  \bibinfo{author}{\bibfnamefont{D.}~\bibnamefont{Villamaina}},
  \bibnamefont{and} \bibinfo{author}{\bibfnamefont{A.}~\bibnamefont{Puglisi}},
  \bibinfo{journal}{J. Stat. Mech.}  (\bibinfo{year}{2011}).

\bibitem[{\citenamefont{P\"oschel and Schwager}(2005)}]{PS05}
\bibinfo{author}{\bibfnamefont{T.}~\bibnamefont{P\"oschel}} \bibnamefont{and}
  \bibinfo{author}{\bibfnamefont{T.}~\bibnamefont{Schwager}},
  \emph{\bibinfo{title}{Computational granular dynamics: models and
  algorithms}} (\bibinfo{publisher}{Springer-Verlag, Berlin},
  \bibinfo{year}{2005}).

\bibitem[{\citenamefont{Montanero and Santos}(2000)}]{MS00}
\bibinfo{author}{\bibfnamefont{J.~M.} \bibnamefont{Montanero}}
  \bibnamefont{and} \bibinfo{author}{\bibfnamefont{A.}~\bibnamefont{Santos}},
  \bibinfo{journal}{Gran. Matt.} \textbf{\bibinfo{volume}{2}},
  \bibinfo{pages}{53} (\bibinfo{year}{2000}).

\bibitem[{\citenamefont{Garz\'o et~al.}(2007)\citenamefont{Garz\'o, Santos, and
  Montanero}}]{GVM07}
\bibinfo{author}{\bibfnamefont{V.}~\bibnamefont{Garz\'o}},
  \bibinfo{author}{\bibfnamefont{A.}~\bibnamefont{Santos}}, \bibnamefont{and}
  \bibinfo{author}{\bibfnamefont{J.~M.} \bibnamefont{Montanero}},
  \bibinfo{journal}{Physica A} \textbf{\bibinfo{volume}{376}},
  \bibinfo{pages}{94} (\bibinfo{year}{2007}).

\bibitem[{\citenamefont{{Vega Reyes}
  et~al.}(2013{\natexlab{a}})\citenamefont{{Vega Reyes}, Santos, and
  Garz\'o}}]{VSG13}
\bibinfo{author}{\bibfnamefont{F.}~\bibnamefont{{Vega Reyes}}},
  \bibinfo{author}{\bibfnamefont{A.}~\bibnamefont{Santos}}, \bibnamefont{and}
  \bibinfo{author}{\bibfnamefont{V.}~\bibnamefont{Garz\'o}},
  \bibinfo{journal}{J. Fluid Mech.} \textbf{\bibinfo{volume}{719}},
  \bibinfo{pages}{431} (\bibinfo{year}{2013}{\natexlab{a}}).

\bibitem[{\citenamefont{{Vega Reyes}
  et~al.}(2013{\natexlab{b}})\citenamefont{{Vega Reyes}, Garz\'o, and
  Khalil}}]{VGK13}
\bibinfo{author}{\bibfnamefont{F.}~\bibnamefont{{Vega Reyes}}},
  \bibinfo{author}{\bibfnamefont{V.}~\bibnamefont{Garz\'o}}, \bibnamefont{and}
  \bibinfo{author}{\bibfnamefont{N.}~\bibnamefont{Khalil}}
  (\bibinfo{year}{2013}{\natexlab{b}}), \bibinfo{note}{preprint in
  arxiv:1310.0771}.

\end{thebibliography}

\end{document}